\documentclass{article}
\usepackage{comment}
\usepackage{graphicx}
\usepackage{cmbright}
\usepackage{optidef}
\usepackage{amsmath}
\usepackage{mathtools}
\usepackage{amssymb}
\usepackage{xcolor}
\usepackage{algorithm}
\usepackage{algpseudocode}
\usepackage{enumitem}

\usepackage{caption}
\usepackage{subcaption}

\usepackage[
doi=false,isbn=false,url=false,eprint=false,
backend=biber,
style=numeric,
sorting=none,
giveninits=true,
maxbibnames=3
]{biblatex}
\renewbibmacro{in:}{}
\AtEveryBibitem{%
  \clearlist{language}%
  \clearfield{pages}
  \clearfield{note}
  \clearfield{month}
}
\title{Reducing the set of considered scenarios in robust optimization of intensity-modulated proton therapy}
\author{Ivar Bengtsson}
\date{April 16, 2025}

\begin{document}

\maketitle

\begin{abstract}

    \noindent Robust optimization is a commonly employed method to mitigate uncertainties in the planning of \textit{intensity-modulated proton therapy} (IMPT). In certain contexts, the large number of uncertainty scenarios makes the robust problem impractically expensive to solve. Recent developments in research on IMPT treatment planning indicate that the number of ideally considered error scenarios may continue to increase.

    In this paper, we therefore investigate methods that reduce the size of the scenario set considered during the robust optimization. Six cases of patients with non-small cell lung cancer are considered. First, we investigate the existence of an optimal subset of scenarios that needs to be considered during robust optimization, and perform experiments to see if the set can be found in a reasonable time and substitute for the full set of scenarios during robust IMPT optimization. We then consider heuristic methods to estimate this subset or find subsets with similar properties. Specifically, we select a subset of maximal diversity in terms of scenario-specific features such as the dose distributions and function gradients at the initial point. Finally, we consider adversarial methods as an alternative to solving the full robust problem and investigate the impact on computation times.

    The results indicate that the optimal subset can be well approximated by solving the robust IMPT problem with conventional methods. Of the methods designed to approximate it within a practically useful time frame, the results of the diversity-maximization methods indicate that they may perform better than a manual selection of scenarios based on the patient geometry. In addition, the adversarial approaches decreased the computation time by at least half compared to the conventional approach.
    
\end{abstract}

\section{Introduction}

\textit{Intensity-modulated proton therapy} (IMPT) allows for the delivery of highly conformal dose distributions thanks to the Bragg peak effect. However, the Bragg peak positions are very sensitive to the density along the beam path, making IMPT sensitive to modeling errors. Robust IMPT optimization is commonly used to mitigate this uncertainty \cite{unkelbach_robust_2018}. A dominating form of robust IMPT planning is the \textit{minimax} approach presented by Fredriksson, which considers a discrete set of scenarios over which to minimize the objective in the worst outcome \cite{fredriksson_minimax_2011}. The most commonly considered error sources are setup uncertainty, related to the positioning of the patient relative to the treatment isocenter, and density uncertainty, which arises when converting CT HU values to proton stopping power.

Another uncertainty source is patient motion, which is a concern when treating tumors in sites such as lung and liver. \textit{4D-robust optimization} (4DRO) is a clinically available mitigation strategy in which each scenario also specifies the image representing the patient's anatomy \cite{janson_treatment_2024}. Typically, the images are taken as the motion phases of a 4DCT. 4DRO is the primary example of robust optimization considering multiple images; another example is anatomically robust optimization \cite{smolders_diffusert_2024}.

A downside of using multiple images is that the number of scenarios considered in the minimax approach increases exponentially with the number of uncertainty sources. For example, 4DRO considering the 10 phases of a 4DCT, 7 setup errors, and three density errors yields 210 scenarios for which to compute dose-influence matrices and include in the numerical optimization. This computational burden challenges widespread clinical implementation \cite{knausl_review_2024}. In addition, with increased modeling capabilities and demands arising in adaptive radiation therapy, there are many new uncertainty sources that may be desirable to incorporate in robust optimization. Examples of such include contour uncertainty and motion uncertainty considering interplay effects \cite{smolders_robust_2024, engwall_4d_2018}.

With an increasing computational burden of robust optimization comes an increased need for efficient optimization algorithms. A central question in this study is whether the scenario set used in robust optimization is needed in full or if the true problem can be approximated by a smaller one that only considers a subset of the scenarios. Previously, Buti et al.\ have proposed an optimization method that only simultaneously considers a subset of the scenarios \cite{buti_accelerated_2020}. The subset is selected heuristically based on a random sampling for which the probability distribution is updated during the optimization process. Their study reports optimization time reductions of 84 \% but is delimited to the case of numerical optimization based on gradient descent. In this study, we consider scenario-selection methods based on a priori maximization of the diversity of the subset and adversarial methods that iteratively include additional scenarios during the optimization process. We then incorporate them in a \textit{sequential quadratic programming} (SQP) algorithm where the scenarios are represented as non-linear constraints. Previously, adversarial methods have been explored for motion-robust radiation therapy planning in the linear setting by Mahmoudzadeh et al.\ \cite{mahmoudzadeh_constraint_2016}.

\section{Mathematical preliminaries}

Consider the minimax formulation of a robust radiation therapy optimization problem given a scenario set $\mathcal{S}$:

\begin{mini}|l|
{x \in \mathcal{X}}{h(d(x; s_1)) + \underset{s \in \mathcal{S}}{\text{max} \quad} f(d(x; s); s).}{}{}
\label{robust_impt}
\end{mini}

\noindent Here, the variable $x$ is constrained to the feasible set $\mathcal{X}$, and $h(\cdot)$ and $f(\cdot)$ are the nominal (with nominal scenario $s_1 \in \mathcal{S}$) and robust objectives, respectively. Both objectives take the dose vector $d(x; s)$ as their argument, but the robust objective may also depend on the scenario in ways other than the dose. Introducing the auxiliary variable $z$, we may equivalently write:

\begin{mini*}|l|
{x \in \mathcal{X}, z \in \mathbb{R}}{h(d(x; s_1)) + z}{}{}
\addConstraint{z}{\geq f(d(x; s); s)}{\quad \forall s \in \mathcal{S}.}\tag{$P_{\mathcal{S}}$}
\label{eq_problem}
\end{mini*}

\noindent We are interested in the case when \textit{full} scenario set $\mathcal{S}$ has too many scenarios for the solution to the minimax problem to be computed in an acceptable time.

At an optimal solution $(x^{\star}_{\mathcal{S}}, z^{\star}_{\mathcal{S}})$ to Problem \eqref{eq_problem} there is an active set of constraints $\mathcal{A}^{\star} \subset \mathcal{S}$, for which $s \in \mathcal{A}^\star \iff z^{\star}_{\mathcal{S}} = f(d(x^{\star}_{\mathcal{S}}; s); s)$. All other constraints are redundant, and Problem \eqref{eq_problem} could be equivalently reformulated without them. Consequently, successfully approximating $\mathcal{A}^{\star}$ in advance would allow for a smaller computational effort required to solve Problem \eqref{eq_problem}.

For any subset $\mathcal{M} \subset \mathcal{S}$, we use $x^{\star}_{\mathcal{M}}$ to denote an optimal solution to the relaxation $(P_{\mathcal{M}})$ of Problem \eqref{eq_problem}, where $\mathcal{M}$ substitutes for $\mathcal{S}$. Also, denote the objective in Problem \eqref{robust_impt} as $F(x; \mathcal{S})$. It then holds that:

\begin{equation}
    F(x^{\star}_{\mathcal{M}}; \mathcal{M}) \leq F(x^{\star}_{\mathcal{S}}; \mathcal{S}) \leq F(x^{\star}_{\mathcal{M}}; \mathcal{S}).
\end{equation}

\noindent Of course, for any feasible $x$, it also holds that

\begin{equation}
    F(x; \mathcal{M}) \leq F(x; \mathcal{S}).
\end{equation}

\noindent It is useful to define the non-negative \textit{robustness gap}, $\Delta (x, \mathcal{M})$, of the pair $(x, \mathcal{M})$ as $\Delta (x, \mathcal{M}) \coloneqq F(x; \mathcal{S}) - F(x; \mathcal{M}).$ With inputs $(x^\star_{\mathcal{M}}, \mathcal{M})$, it quantifies the increase in objective value as the full scenario set is evaluated after having optimized with respect to $\mathcal{M}$.

\section{Purpose}

Much of the theory above depends on all involved optimization problems being solved exactly. In radiation therapy treatment planning, however, it is common to formulate problems with non-linear and non-smooth objective and constraint functions, such as sums of squared one-sided voxel-wise dose deviations, for which initial progress towards creating a good plan is made quickly, but convergence toward the exact optimal solution is considerably slower and typically not prioritized. Instead, the user may choose to terminate the optimization process whenever the resulting dose is good enough or when the current problem formulation is deemed unsuitable. Therefore, there are several interesting questions about the mathematical theory's applicability to solving radiation therapy problems in practice. We divide them into two groups:

\begin{enumerate}
    \item First, is the exact optimal subset $\mathcal{A}^{\star}$ \textit{useful} in practice, in the sense that it gives a low $\Delta (x^\varepsilon_{\mathcal{A}^{\star}}, \mathcal{A}^{\star})$ where $x^\varepsilon_{\mathcal{A}^{\star}}$ is an approximate solution to problem $(P_{\mathcal{A}^\star})$? In essence, this means that the objective value does not deteriorate much when considering the full set of scenarios after using the optimal subset during optimization, even if the numerical optimization is terminated early.
    
    If so, how does the usefulness (in the above sense) of an approximation $\mathcal{A}^{\varepsilon}$ of $\mathcal{A}^{\star}$, based on an approximate solution $x^\varepsilon_\mathcal{S}$, depend on the tolerance specified by $\varepsilon$? Can $\mathcal{A}^{\star}$ (or a small superset thereof) be reliably found by identifying the scenarios with worst objective values for reasonable values of $\varepsilon$?

    \item If the answer to these questions is that $\mathcal{A}^{\star}$ itself and its estimates $\mathcal{A}^\varepsilon$ are all useful, we are left with the problem that our method for approximating $\mathcal{A}^{\star}$ is based on solving Problem \eqref{eq_problem}. For drawing any practical value from the usefulness of $\mathcal{A}^{\star}$, it is of interest whether there are other ways to estimate it or if other methods can be used to specify subsets that also have small robustness gaps. Is it, for example, possible to use the problem parameters to determine which scenarios will, if taken as the subset, result in a low robustness gap?
    
\end{enumerate}

\section{Scenario subset selection} \label{scenario_selection}

This section revolves around the intuitive idea that $\mathcal{A}^{\star}$, and good approximations thereof, contain the scenarios whose corresponding functions are in most conflict with each other and with the nominal objective function. Hence, we wish to select a small subset of maximally different scenarios, preserving as much of the variation in $\mathcal{S}$ as possible.

\subsection{Manual selection of extreme scenarios}

It may be possible to select a good subset $\mathcal{M}$ simply by considering the patient geometry. For example, when considering setup shifts within a zero-centered sphere with radius $r$, selecting setup shifts of orthogonal directions and magnitude $r$ may be sufficiently robust to errors within the whole sphere. Likewise, in the 4D-robustness context, selecting 4DCT phases at the two extremes of the breathing cycle may suffice.

\subsection{Distance-based selection}

Beyond manual selection, it may be useful to select a diverse set of scenarios based on some computed distances between pairs of scenarios.

\subsubsection{Dose-based differences}

Arguably, the most important difference between any two scenarios $s$ and $s'$ is that between their impacts on the dose. Thus, a natural space in which to measure the distance between scenarios is that of their dose-influence matrices $D_s, \forall s \in \mathcal{S}$. We could, for example, define the distance by

\begin{equation}
    \delta^{\text{dose-influence}}(s, s') \coloneq \| D_s - D_{s'} \|
\end{equation}

\noindent for some choice of matrix norm. It may, however, be of interest to reduce the dimensionality of the space by multiplying each dose-influence matrix by some $x_{\delta}$ to compute a scenario dose distribution, $d(x_{\delta}; s) = D_s x_{\delta}$. Using the euclidean norm, the distance computation is then:

\begin{equation}
    \delta^{\text{dose}}(s, s') \coloneq \| d(x_{\delta}; s) - d(x_{\delta}; s') \|_2.
    \label{dose_distance}
\end{equation}

\subsubsection{Gradient-based differences}

Another option, which, besides the dose distributions, also accounts for the objective function and the ROI geometries, is to base the distances on the scenario function gradients. We could, for example, consider the negative of the cosine similarity that measures the angle between the gradient directions:

\begin{equation}
    \delta^{\text{grad}}(s, s') \coloneq \frac{\nabla f(d(x_{\delta}, s); s)^T \nabla f(d(x_{\delta}, s'); s')}{\| \nabla f(d(x_{\delta}, s); s) \| _2 \cdot \| \nabla f(d(x_{\delta}, s'); s') \|_2}.
    \label{gradient_distance}
\end{equation}

\begin{comment}

\end{comment}

\subsection{Maximizing diversity}

The final component of distance-based selection is the maximization of the diversity in terms of the computed distances, given a maximum cardinality of $K$. There are many plausible formulations. We may, for example, want to solve the problem:

\begin{maxi}|l|
{\mathcal{M} \in \mathcal{S}}{\underset{i,j \in \mathcal{M}}{\text{min} \quad} \delta(s_i, s_j)}{}{}
\addConstraint{|\mathcal{M}|}{= K}{,}
\end{maxi}

\noindent for any of the above definitions of $\delta(s, s')$.

\section{Building the scenario set adversarially} \label{adversarial_approach}

\noindent Because the approaches outlined above do not guarantee to find an $\mathcal{M}$ with a low robustness gap, we may also want to add scenarios as the numerical optimization progresses. Consider the partition of $\mathcal{S}$ into an active set 
$\mathcal{A}$ and an inactive set $\mathcal{I}$, satisfying $\mathcal{A} \cup \mathcal{I} = \mathcal{S}, \mathcal{A} \cap \mathcal{I} = \emptyset$. The procedure presented in Algorithm \ref{alg_adversarial}, following Mutapcic \& Boyd \cite{mutapcic_cutting-set_2009}, finds a solution that is suboptimal by at most $\varepsilon$ for Problem \eqref{eq_problem}.

\begin{algorithm}
\caption{Adversarial minimax}
\begin{algorithmic}[1]
    \State Initialize $\mathcal{A} = \mathcal{M}$, $\mathcal{I} = \mathcal{S} \setminus \mathcal{A}$.
    \State Solve $(P_{\mathcal{A}})$ to obtain the approximate solution $(x^\star_{\mathcal{A}}, z^\star_{\mathcal{A}})$.
    \State Identify the worst scenario in $\mathcal{I}$ as the solution $s^\star$ to 
    \begin{maxi*}|l|
    {s \in \mathcal{I}}{f(d(x^\star_{\mathcal{A}}; s); s) - z^\star_{\mathcal{A}}.}{}{}
    \end{maxi*}
    
    \If{$\Delta (x^\star_{\mathcal{A}}, \mathcal{A}) \leq \varepsilon $}
        \State Exit
    \Else:
        \State Update the sets as $\mathcal{A} \leftarrow \mathcal{A} \cup \{ s^\star \}$, $\mathcal{I} \leftarrow \mathcal{I} \setminus \{ s^\star \}$, and go to step 2.
    \EndIf
        
\end{algorithmic}
\label{alg_adversarial}
\end{algorithm}

\noindent In the next section, we present modifications to Algorithm \ref{alg_adversarial} that were made for the numerical experiments.

\section{Numerical experiments}

\subsection{Patient cases}

We considered data from six patients with non-small cell lung cancer, uploaded to The Cancer Imaging Archive \cite{clark_cancer_2013}. For each patient, there was a 4DCT with the CTV contoured on every motion phase and various OAR contours on one or more phases. The selected patients exhibit a variation of tumor size and motion magnitude. The full dataset is described in detail in Hugo et al.\ \cite{hugo_longitudinal_2017}.

For each patient, a two-beam IMPT plan was created in RayStation (RaySearch Laboratories AB, Stockholm, Sweden). The OARs were the heart, the spinal cord, and the esophagus. The weights of their dose-limiting maximum dose functions were kept constant across all patients, while the dose levels were adjusted to force relevant optimization trade-offs with the target objectives. In addition, a maximum dose objective was for all patients applied to the lung (on the side of the tumor) minus a 1 cm expansion of the CTV (CTV+1cm). The target objectives were a minimum dose objective for the CTV, with the dose level set to a prescribed dose of 66 Gy, and a maximum dose objective with dose level 69.3 Gy (105\% of the prescribed dose) to the CTV+1cm.

The considered uncertainty scenarios were the 210 combinations of the ten motion phases of the 4DCT, setup shifts of 0 mm and 5 mm in each of the six axial directions, and density uncertainties of 0 and $\pm 3.5 \%$. After finalizing the optimization problem parameters, the ROI geometries and the dose-influence matrices on a voxel-grid of $5 \times 5 \times 5$ mm were exported to Matlab (2024a, Mathworks Inc, Natick, Massachusetts, USA). There, the optimization problems were solved using the API to the SQP algorithm SNOPT (7.7, Stanford Business Software, Stanford, California). For all optimizations, the initial point $x_0$ was taken as the initial guess from RayStation.

\subsection{Experiment 1: Existence of $\mathcal{A}^\star$}

The mathematical preliminaries state that $\mathcal{A}^\star$ exists. However, they do not indicate if an SQP method used to solve radiation therapy problems will find a sufficiently accurate solution to allow for the determination of $\mathcal{A}^\star$ in reasonable time. Therefore, we here present the method used to estimate $\mathcal{A}^\star$.

For each patient case, we first solved Problem \eqref{eq_problem} with SNOPT tolerances equal to $\varepsilon = 10^{-5}, \varepsilon = 10^{-6}$, and $\varepsilon = 10^{-7}$. Each solution was denoted $(x^{\varepsilon}_\mathcal{S}, z^{\varepsilon}_\mathcal{S})$. Then, the estimates $\mathcal{A}^\varepsilon$ of $\mathcal{A}^\star$ was defined as

\begin{equation}
 \mathcal{A}^\varepsilon \coloneqq \{ s \in \mathcal{S} : f(d(x^{\varepsilon}_\mathcal{S}; s); s) > z_\mathcal{S}^{\varepsilon} - \tau \},  
\end{equation}
 
\noindent where the threshold value $\tau$ was primarily selected as 

\begin{equation}
    \tau = \frac{1}{100}[ \underset{s \in \mathcal{S}}{\text{max} \quad} f(d(x^{\varepsilon}; s); s) - \underset{s \in \mathcal{S}}{\text{min} \quad} f(d(x^{\varepsilon}; s); s) ].
    \label{tau}
\end{equation}

\noindent Finally, $\mathcal{A}^\varepsilon$ was verified by solving Problem $(P_{\mathcal{A}^\varepsilon})$, obtaining the solution $x_{\mathcal{A}^\varepsilon}^\varepsilon$, and comparing the objective values of $x^\varepsilon_{\mathcal{S}}$ and $x^\varepsilon_{\mathcal{A}^\varepsilon}$. Similar objective values indicate the usefulness of $\mathcal{A}^\varepsilon$, as it means that $\mathcal{A}^\varepsilon$ can substitute for $\mathcal{S}$ in optimization without substantial consequences for the robustness of the resulting solution.

\subsection{Experiment 2: Usefulness of $\mathcal{A}^\star$ and its estimates}

Experiment 2 was built on experiment 1 in the sense that we investigated whether the robustness gap of $\mathcal{A}^\varepsilon$ was good also earlier during the numerical optimization to determine the usefulness of $A^\varepsilon$ under normal radiation therapy treatment planning circumstances, where optimization problems are rarely solved to optimality. To this end, for any estimate $\mathcal{M}$ of $\mathcal{A}^\star$ (including $\mathcal{M} = \mathcal{A}^\varepsilon$ with $\varepsilon = 10^{-7}$), we computed, for $k = 50$, the true objective function value $F(x^k_{\mathcal{M}}; \mathcal{S})$, where $x^k_{\mathcal{M}}$ denotes the $k$:th iterate of the optimization trajectory when solving problem $(P_{\mathcal{M}})$.

The remaining methods to select $\mathcal{M}$ were based on the principles from Section \ref{scenario_selection}. As benchmark, a manual selection of 24 \textit{extreme} scenarios was made based on our intuition, considering the six non-zero setup shifts, the non-zero density scenarios, and two motion phases: the reference phase (end-exhale) and the phase with the largest CTV displacement (as measured by the mean magnitude in the CTV of the displacement vector field from the deformable image registration aligning each of the non-reference phases with the reference phase). The set of these 24 scenarios is denoted by $\mathcal{E}$.

Then, two methods selecting 24 scenarios by maximizing the diversity of scenario-specific features were implemented. The first was based on dose distributions at the initial point and measured pairwise distances as in Equation \eqref{dose_distance}, with $x_\delta = x_0$. The second used the gradient-based distances described in Equation \eqref{gradient_distance}, measured in the same point. For both diversity-maximization methods, a greedy heuristic was implemented to select a subset with maximal minimal distance between any two elements. After the greedy selection, swaps of a selected scenario with a non-selected scenario were made until no available swaps improved the diversity objective. In the dose-based method, the first scenario was that which was maximally different from the nominal scenario. In the gradient-based method, it was the one maximally different from the nominal objective gradient, $\nabla_x h(d(x_0; s_1))$. Furthermore, this nominal gradient was then consistently included when computing the diversity objective to maintain a scenario selection in conflict not only internally but also with the nominal objective. The resulting sets are denoted by $\mathcal{D}$ and $\mathcal{G}$, respectively.

\subsection{Experiment 3: Adversarial methods}

Finally, we wished to investigate whether any computational gains could be made by the subset selection methods from Section \ref{scenario_selection} or the adversarial approach from Section \ref{adversarial_approach}. To this end, we compared the optimization progress of seven methods. The benchmark method was to directly model and solve the full Problem \eqref{eq_problem} with SNOPT, as finding $\mathcal{A}^\star$ according to the method in experiment 1 is equally time-consuming and offers no computational benefit.

The six remaining methods were adversarial. The first three used only the nominal scenario as the initial subset $\mathcal{M}$ in Algorithm \ref{alg_adversarial}. They differed by their strategy of appending scenarios in step 3:

\begin{itemize}
    \item The first method added the inactive scenarios with the worst constraint violations $f(d(x^\star_{\mathcal{A}}; s); s) - z^\star_{\mathcal{A}}$, among those with more than 1 \% of the maximal violation.
\end{itemize}

\noindent The second and third methods selected scenarios from those with more than 25 \% of the maximal violation.

\begin{itemize}[resume]
    \item Out of the available scenarios, the second method added those with the greatest diversity in dose at the current point.
    \item The third method added those with the greatest diversity in gradient at the current point.
\end{itemize}

\noindent In both methods, the diversity-maximization heuristic was initialized from the scenario with maximal constraint violation. In addition, this maximally violated scenario was not considered by the greedy swaps.

The remaining three adversarial methods used the same three principles for scenario addition, but initialized Algorithm \ref{alg_adversarial} with the maximally diverse subsets from experiment 2. All in all, the compared methods are summarized in Table \ref{adversarial_methods}.

\begin{table}[h]
    \centering
    \begin{tabular}{l|l|c}
        \hline
        \textbf{Initial Selection ($\mathcal{M}$)} & \textbf{Method to add scenarios} & \textbf{Short} \\
        \hline
        All scenarios $(\mathcal{S})$ & - & - \\
        Nominal scenario & Worst scenarios by objective & NO \\
        Nominal scenario & Diversity in dose & ND \\
        Nominal scenario & Diversity in gradient & NG \\
        Extreme selection $(\mathcal{E})$ & Worst scenarios by objective & EO \\
        Maximal diversity of doses $(\mathcal{D})$ & Diversity in dose & DD \\
        Maximal diversity of gradients $(\mathcal{G})$ & Diversity in gradient & GG \\
        \hline
    \end{tabular}
    \caption{Summary of the adversarial methods.}
    \label{adversarial_methods}
\end{table}

\noindent Any method was deemed complete after reaching the objective value $F(x^{50}_{\mathcal{S}}; \mathcal{S})$, corresponding to performing 50 SQP iterations on Problem \eqref{eq_problem}. Furthermore, the minimization over the currently active subset $\mathcal{A}$, corresponding to step 2 of Algorithm \ref{alg_adversarial}, was made approximate by stopping the optimization after 8 SQP iterations.

\section{Results}

\subsection{Experiment 1}

\begin{figure}
    \centering
    \makebox[\textwidth][c]{%
        \includegraphics[width=1.2\textwidth]{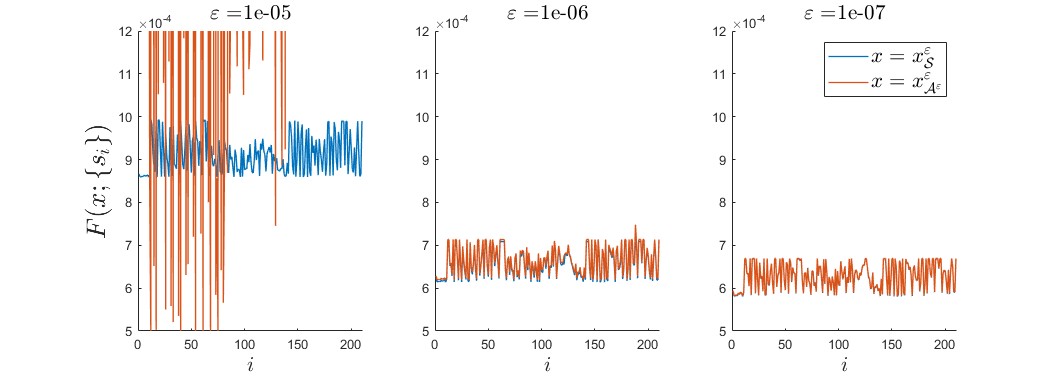}%
    }
    \caption{The \textit{scenario-specific} objective values for P101, for three values of $\varepsilon$ and the corresponding approximate solutions $x^\varepsilon_{\mathcal{S}}$ and $x^\varepsilon_{\mathcal{A}^\varepsilon}$.}
    \label{scenario_values_P101}
\end{figure}

For each case, the number of scenarios in $\mathcal{A}^\varepsilon$ varied with the tolerance $\varepsilon$. These results are presented in  Table \ref{experiment1_table}, alongside the relative differences in objective value ($F(x; \mathcal{S})$) when optimizing considering either the full set $\mathcal{S}$ or the set $\mathcal{A}^\varepsilon$. These differences indicate the usefulness of $\mathcal{A}^\varepsilon$. In addition, Figure \ref{scenario_values_P101} shows for each scenario index $i$ the scenario-specific objectives $F(x; \{ s_i \})$ of P101 for $x = x^\varepsilon_\mathcal{S}$ and $x = x^\varepsilon_{\mathcal{A}^\varepsilon}$, where $F(x; \{ s_i \}) = h(d(x; s_1)) + f(d(x; s_i); s_i)$.

Compared to using the finer tolerances, using $\varepsilon = 10^{-5}$ resulted in small and non-robust sets $\mathcal{A}^\varepsilon$. At this tolerance, the worst scenarios differed too much for the implemented selection procedure to work as intended. As $\varepsilon$ decreases, the similarity between the scenario-specific objectives (including the maximums considered in Table \ref{experiment1_table}) increases. For example, the plotted curves in Figure \ref{scenario_values_P101} become visually indistinguishable. Although not shown graphically, this result is the same across all patients and indicates that it is indeed feasible to sufficiently accurately identify $\mathcal{A}^\star$ by solving IMPT optimization problems with an SQP method. In an attempt to resolve the inaccuracy for $\varepsilon = 10^{-5}$, we additionally considered a more inclusive selection by increasing the threshold $\tau$ by a factor of 10. The resulting values are shown in parenthesis in Table \ref{experiment1_table}.

\begin{table}[h!]
\centering
\makebox[\textwidth]{ % Ensures centering
\begin{tabular}{c|ccc|ccc}
\hline

& \multicolumn{3}{c|}{$|\mathcal{A}^\varepsilon|$} 
& \multicolumn{3}{c}{$\frac{F(x^\varepsilon_{\mathcal{A}^\varepsilon}; \mathcal{S}) - F(x^\varepsilon_\mathcal{S}; \mathcal{S})}{F(x^\varepsilon_\mathcal{S}; \mathcal{S})}$} \\ \cline{2-7}

& \multicolumn{3}{c|}{$\varepsilon$} 
& \multicolumn{3}{c}{$\varepsilon$} \\
& $10^{-5}$ & $10^{-6}$ & $10^{-7}$ 
& $10^{-5}$ & $10^{-6}$ & $10^{-7}$ \\ \hline

P101 & 5 (30) & 39 & 48 & 9756.80\% (6.67\%) & 5.58\% & 0.16\% \\ \hline
P103 & 5 (22) & 39 & 43 & 2301.13\% (-5.74\%) & 3.49\% & 0.03\% \\ \hline
P104 & 4 (47) & 66 & 80 & 1490.17\% (-5.25\%) & 1.27\% & 0.03\% \\ \hline
P110 & 13 (21) & 32 & 33 & 112.37\% (-1.50\%) & -0.46\% & 0.05\% \\ \hline
P111 & 20 (26) & 33 & 33 & 25.41\% (23.45\%) & 0.15\% & 0.02\% \\ \hline
P114 & 2 (11) & 31 & 43 & 1106.40\% (8.18\%) & 2.53\% & 0.48\% \\ \hline
\end{tabular}}
\caption{Size of $\mathcal{A}^\varepsilon$ and relative difference of objective values for different values of the SNOPT tolerance $\varepsilon$. The values in parentheses for $\varepsilon = 10^{-5}$ indicate the result when the definition of $\mathcal{A}^\varepsilon$ was modified by increasing $\tau$, from Equation \eqref{tau}, by a factor of 10.}
\label{experiment1_table}
\end{table}

\subsection{Experiment 2: Usefulness of $\mathcal{A}^\star$ and its estimates}

\begin{figure}
    \centering
    \includegraphics[width=\linewidth]{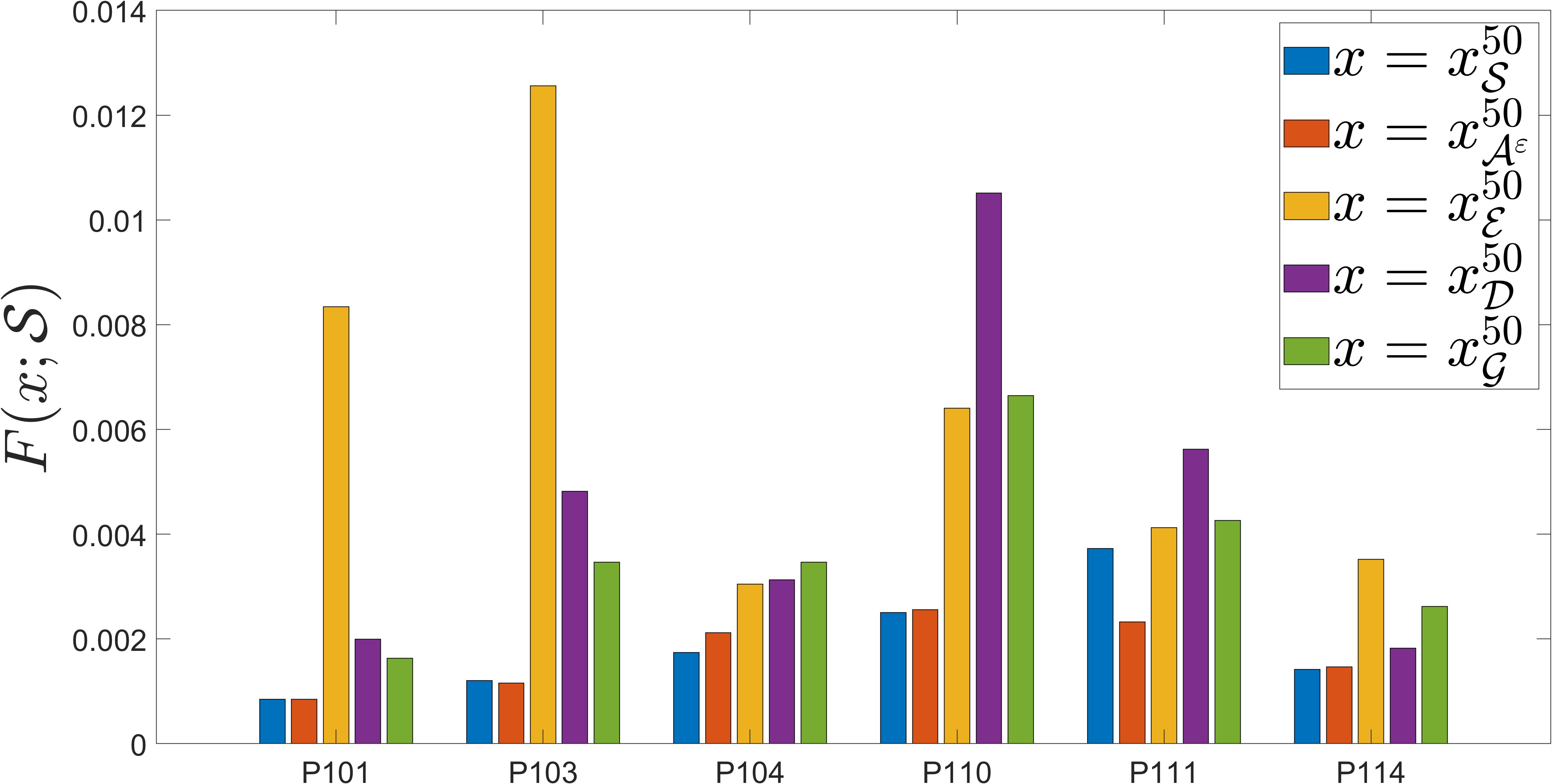}
    \caption{Objective values per subset selection method and case, after 50 SQP iterations. For each compared subset $\mathcal{M}$, the solution after 50 iterations is denoted $x_\mathcal{M}^{50}$. The subset $\mathcal{A}^\varepsilon$ was computed as in experiment 1, with $\varepsilon = 10^{-7}$.}
    \label{selection_usefulness}
\end{figure}

The usefulness (in the sense of resulting in a low objective value $F(x; \mathcal{S})$ after 50 SQP iterations) of each scenario-set selection method is indicated in Figure \ref{selection_usefulness}. For each patient and scenario subset $\mathcal{M}$, the objective value $F(x^{50}_\mathcal{M}; \mathcal{S})$ is indicated by the height of the relevant bar. Compared to using the full scenario-set $\mathcal{S}$, $\mathcal{A}^\varepsilon$ kept the objective value notably low compared to the diversity-maximization methods. In particular, there were two cases, P101 and P103, for which the manually selected extreme scenarios $\mathcal{E}$ failed to compete with the other methods and resulted in many times higher objective values. Similarly, the set $\mathcal{D}$, based on dose diversity, was particularly poor for P110. Gradient-based diversity, with the set denoted by $\mathcal{G}$, was the most consistent among the methods using designed subsets and was marginally worse than the others only for P104.

\subsection{Experiment 3: Adversarial methods}

\begin{figure}
    \centering
    \begin{minipage}{0.48\textwidth}
        \centering
        \includegraphics[width=\linewidth]{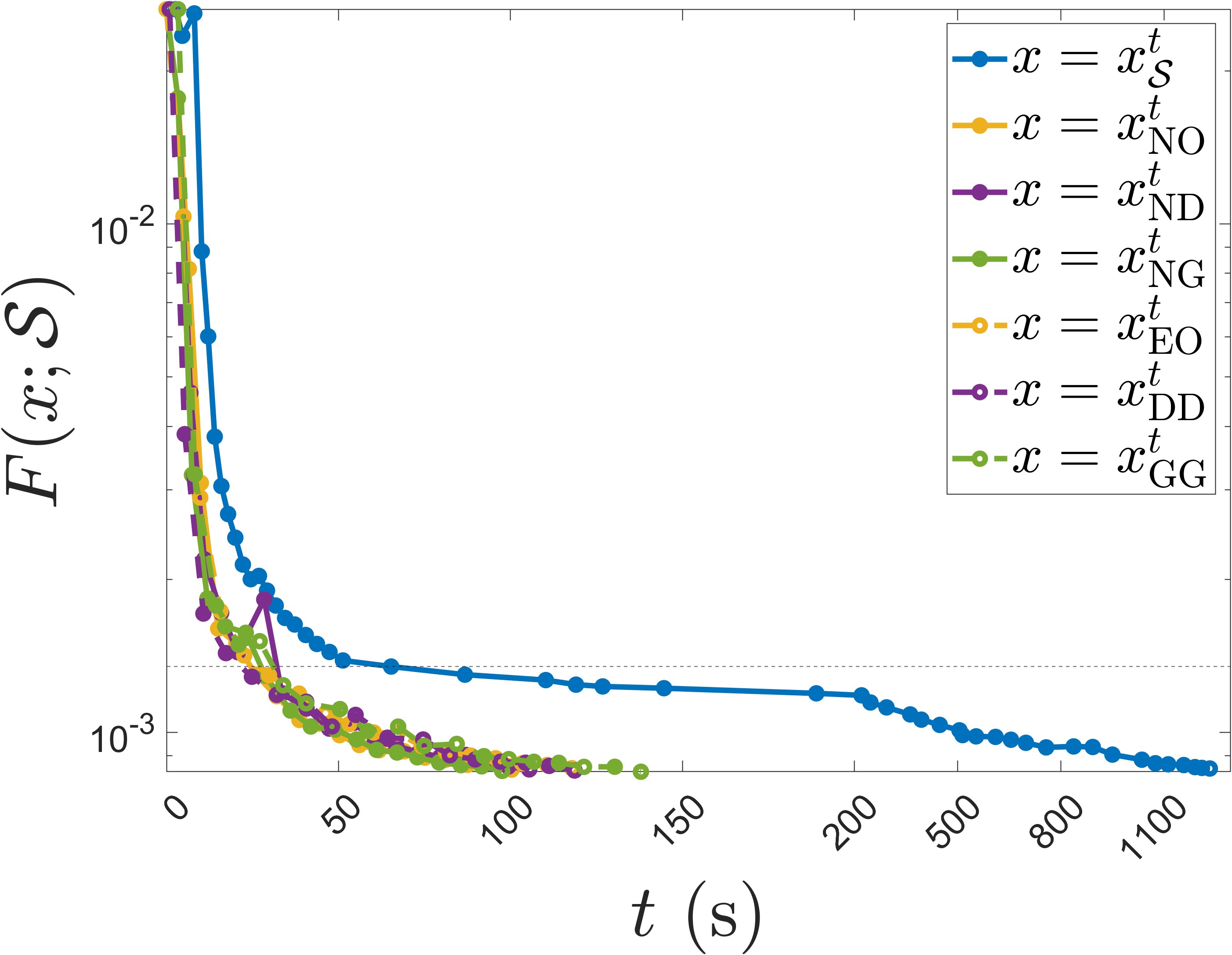}
        \subcaption{P101}
    \end{minipage}
    \hfill
    \begin{minipage}{0.48\textwidth}
        \centering
        \includegraphics[width=\linewidth]{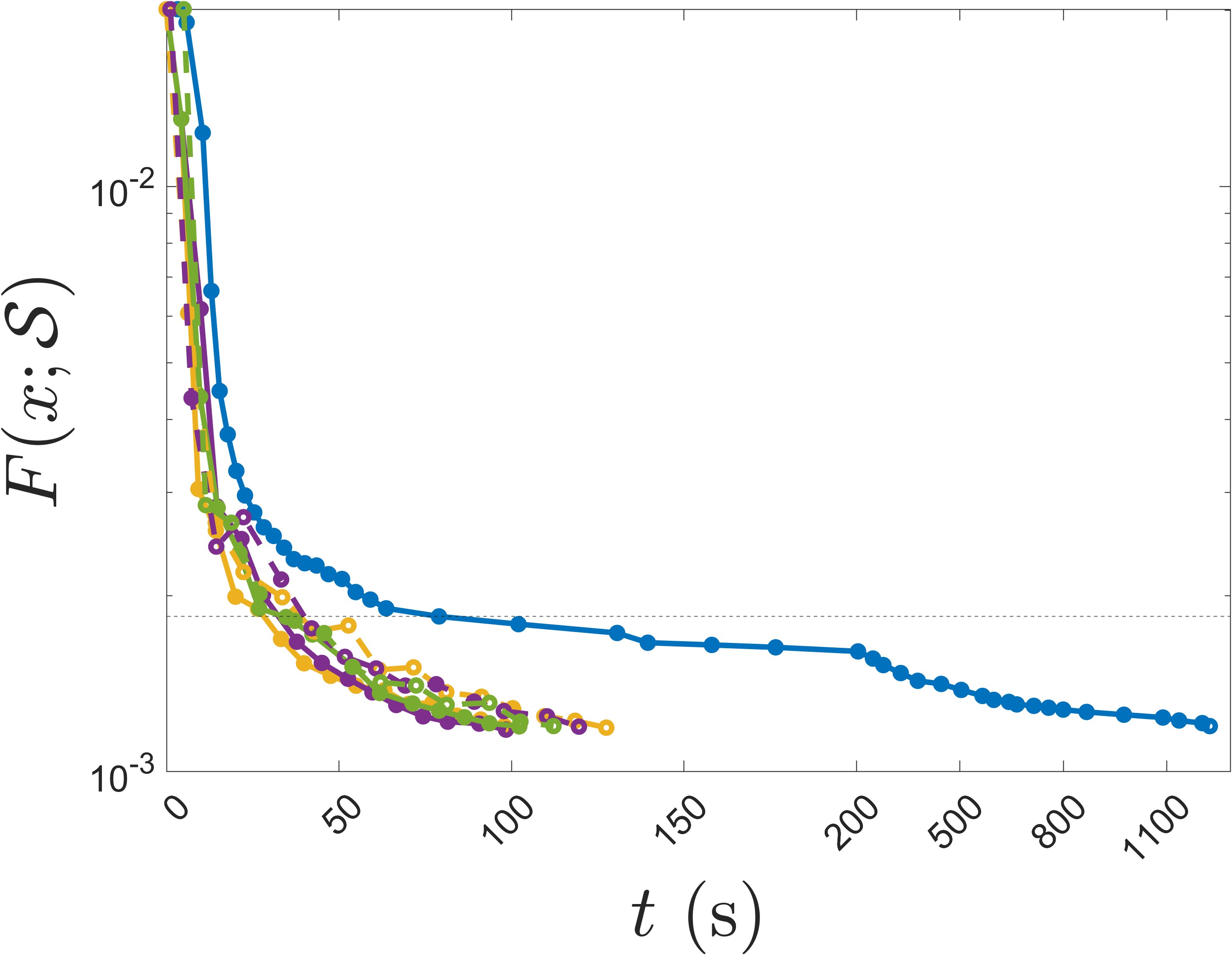}
        \subcaption{P103}
    \end{minipage}
    
    \vspace{0.5cm} % Adjust vertical spacing

    \begin{minipage}{0.48\textwidth}
        \centering
        \includegraphics[width=\linewidth]{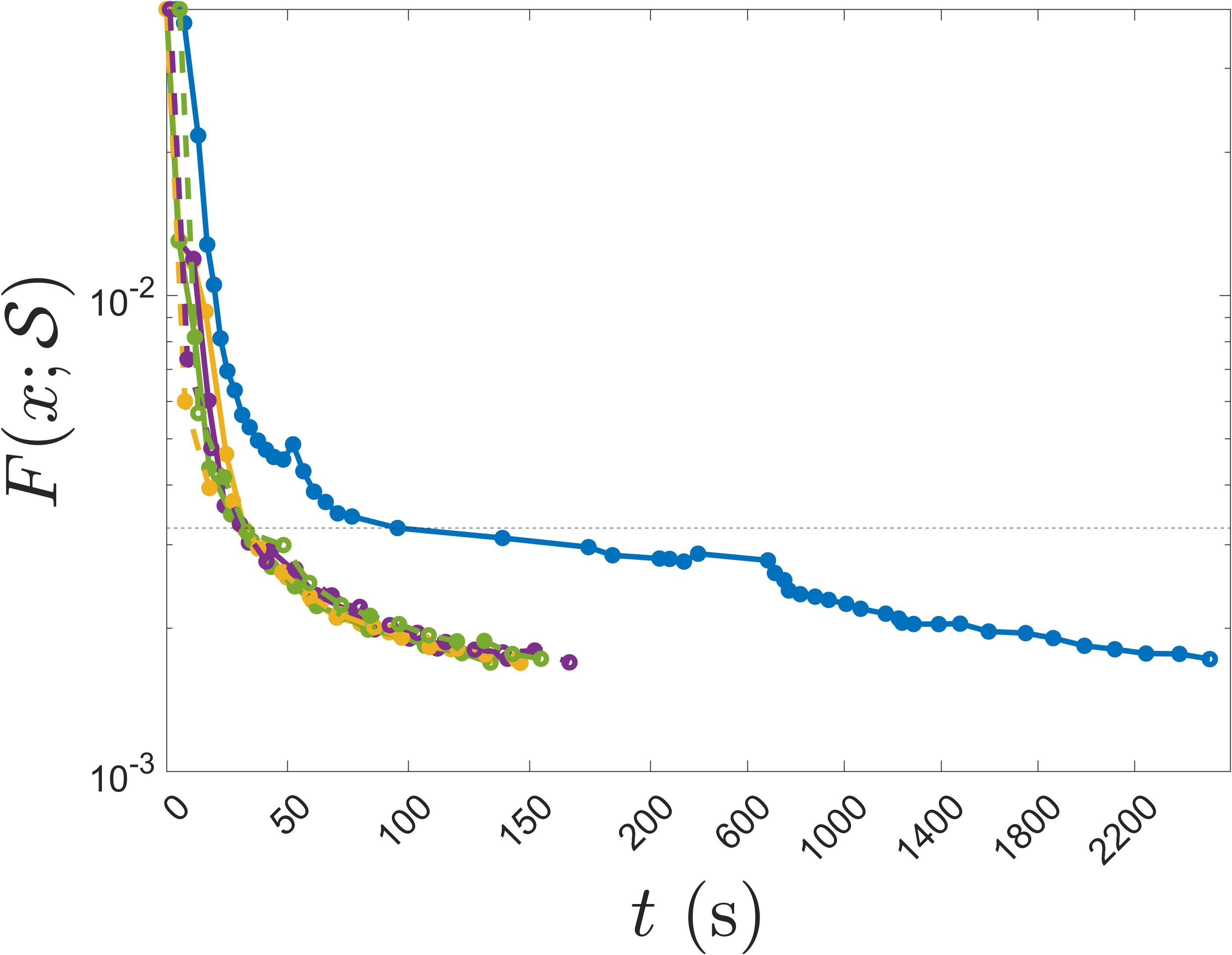}
        \subcaption{P104}
    \end{minipage}
    \hfill
    \begin{minipage}{0.48\textwidth}
        \centering
        \includegraphics[width=\linewidth]{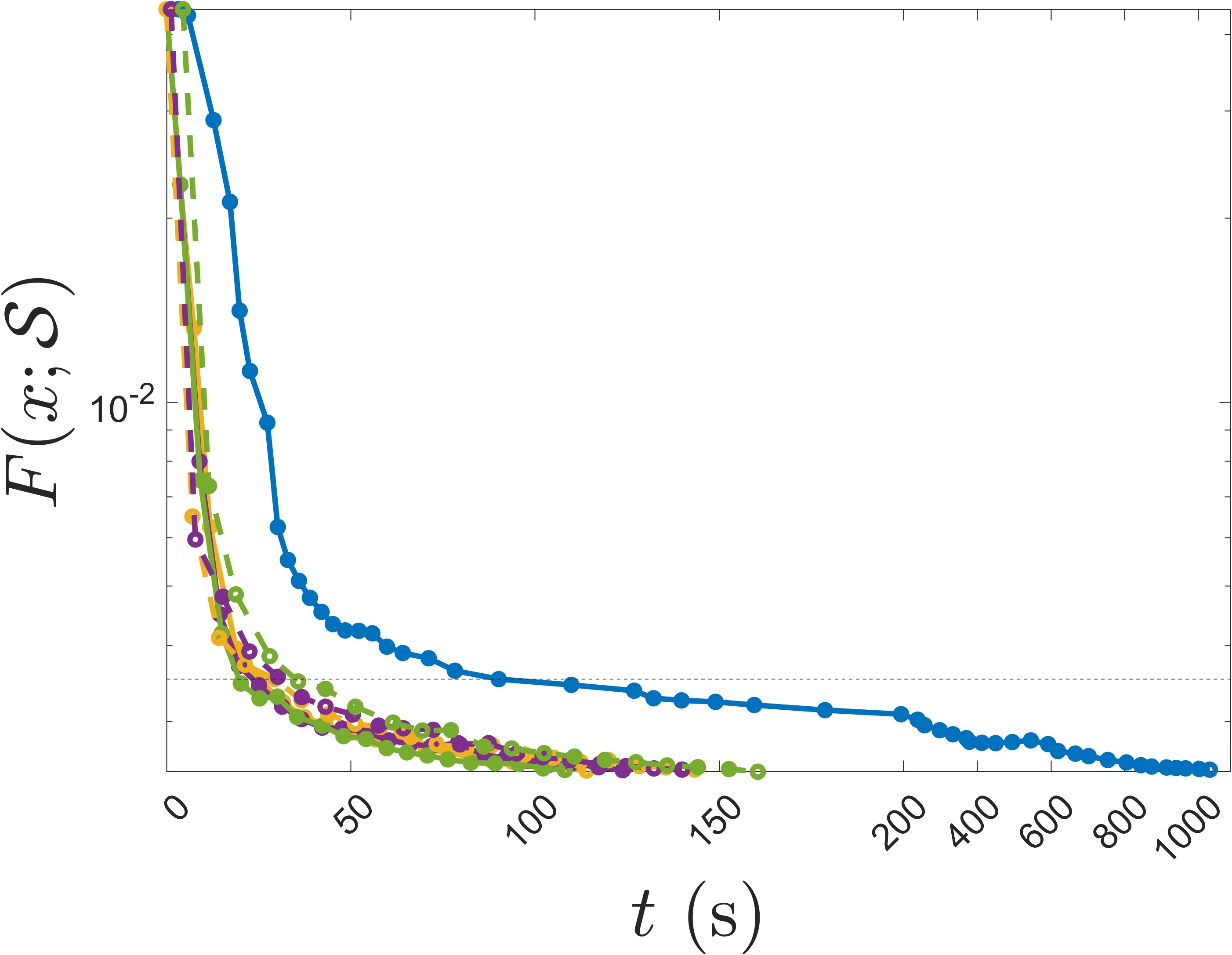}
        \subcaption{P110}
    \end{minipage}

    \vspace{0.5cm} % Adjust vertical spacing

    \begin{minipage}{0.48\textwidth}
        \centering
        \includegraphics[width=\linewidth]{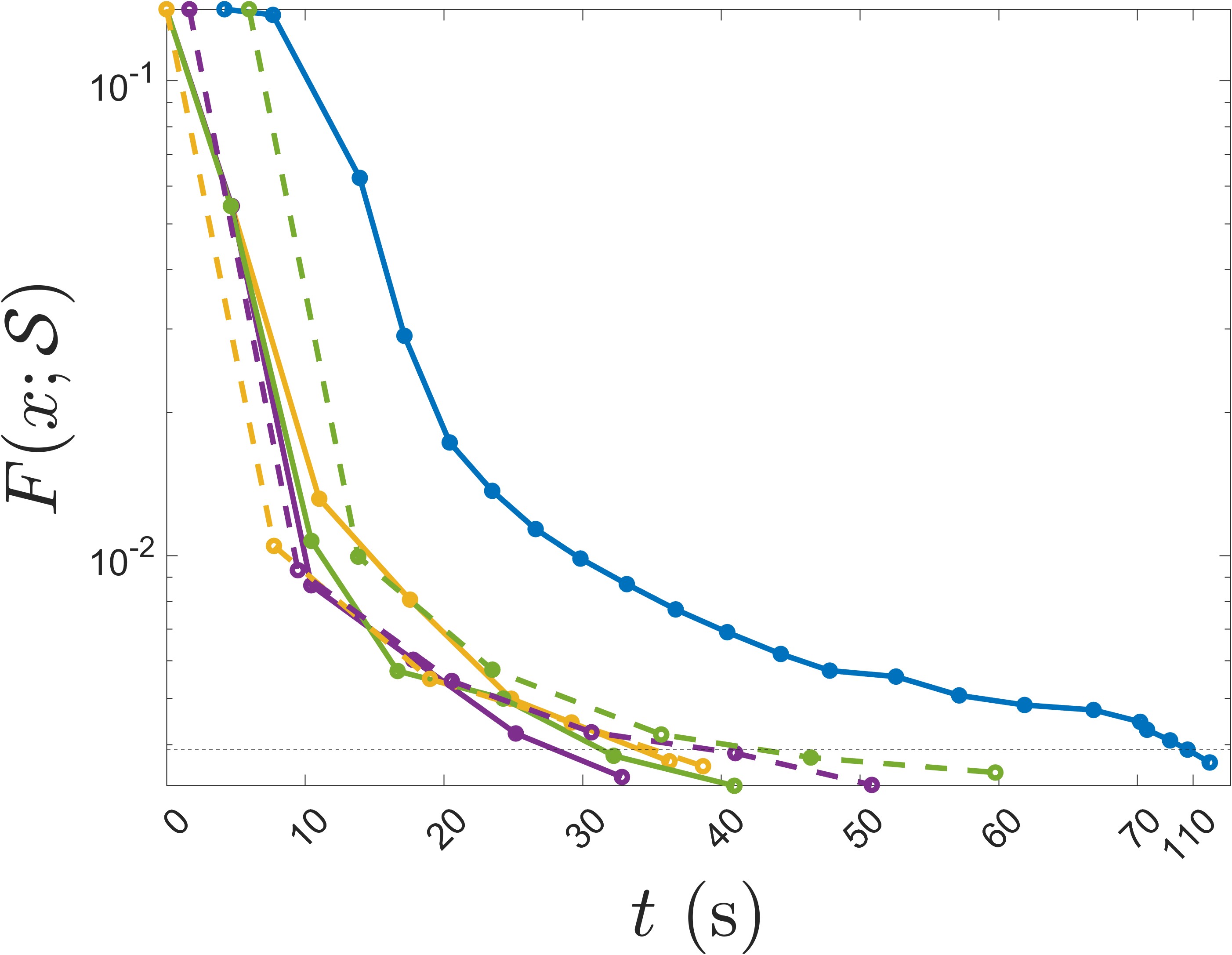}
        \subcaption{P111}
    \end{minipage}
    \hfill
    \begin{minipage}{0.48\textwidth}
        \centering
        \includegraphics[width=\linewidth]{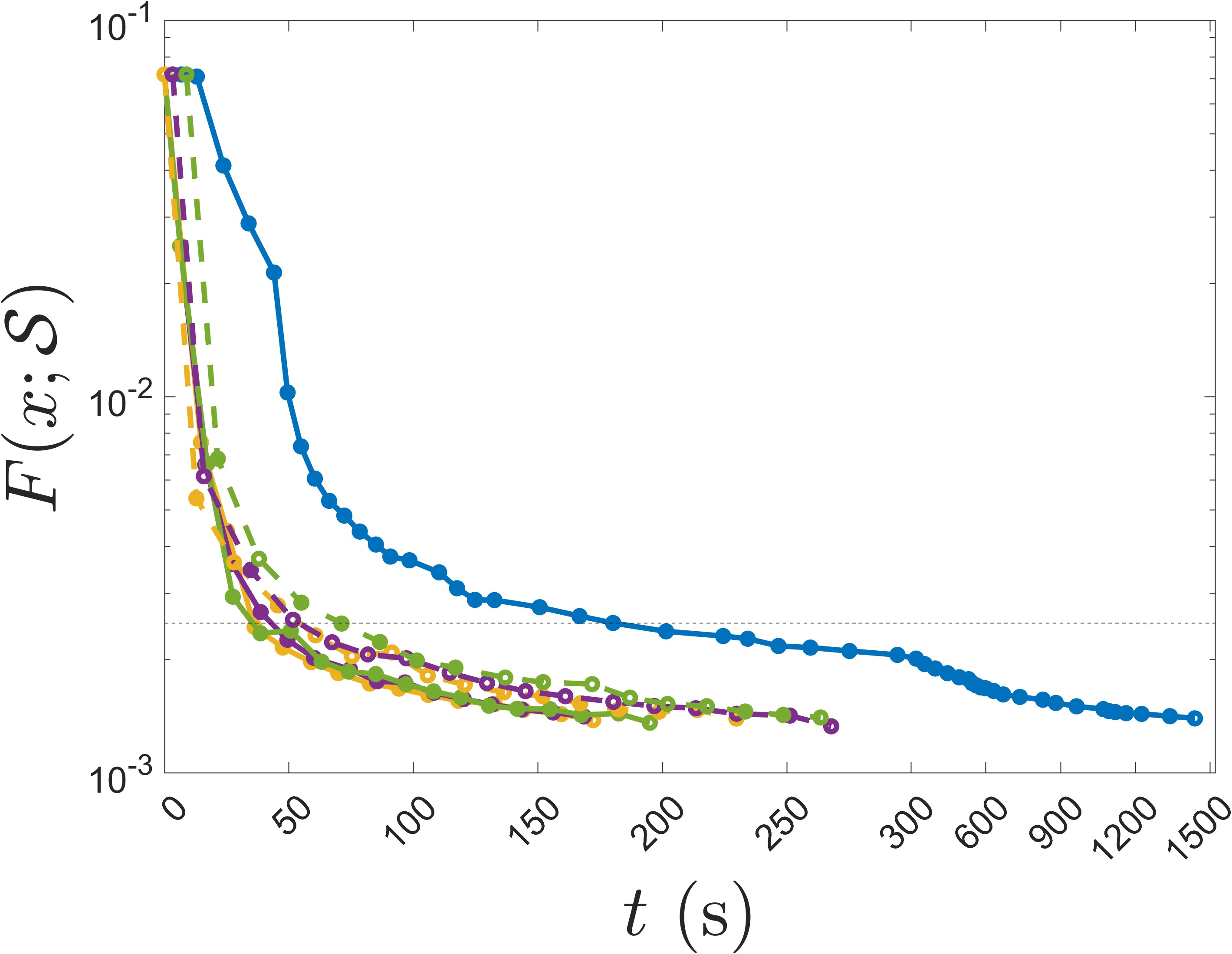}
        \subcaption{P114}
    \end{minipage}
    
    \caption{Optimization progress (objective value over time) of the adversarial methods, compared to that of the benchmark, labeled $x^t_{\mathcal{S}}$, where the superscript now denotes the cumulative computation time $t$. For the adversarial methods, the subscript contains the method's shorthand notation. Note that the objective value is shown on a logarithmic scale while the time axis is shown on a discontinuous linear scale where the right side has been compressed by a factor of 10. The dashed line indicates the objective value of the benchmark after 20 iterations.}
    \label{progress}
\end{figure}

Figure \ref{progress} shows the numerical optimization progress of the adversarial methods, next to that of directly solving Problem \eqref{eq_problem}, acting as the benchmark. The objective value is always the true objective $F(x; \mathcal{S})$, evaluated with the full scenario set. This value was computed only at step 3 of Algorithm \ref{alg_adversarial} for the adversarial methods. For the benchmark, the objective was computed at every major SQP iteration.

Across all cases, the adversarial methods reached the same objective value in less time than the benchmark. In particular, the adversarial methods' relative advantage was greater later in the optimization process, as solving the subproblems in the later iterations of the benchmark method often required considerably more time than during earlier iterations. To reach the objective value $F(x^{50}_\mathcal{S}; \mathcal{S})$, the adversarial methods considerably reduced the required computation time, to (in mean) 5.9--15.0 \% of that of the benchmark for patients P101, P103, P104, P110, and P114, and to 35.4\% for P111.

In comparison, the performance among the adversarial methods was relatively even. The methods initialized with the nominal scenario terminated before their counterparts initiated with a larger scenario set, with few exceptions. There were, however, times earlier during the optimization where those initialized with the nominal scenario were behind those initialized with a larger set, in terms of the value of the objective function.

To nuance the interpretation of the results, we also present differences in the computation times earlier during the optimization, before the slowdown of the benchmark method. For each adversarial method, we recorded the time needed to reach the objective value of the benchmark method after 20 iterations. At this stage, the adversarial methods (in mean) needed less than 40\% of the computation time for P104, P110, P111, and P114 (37.4\%, 33.2\%, 35.7\%, and 29.9\%, respectively), and less than 50\% for P101 and P103 (45.1\% and 48.0\%, respectively).

\section{Discussion}

In this paper, we have described mathematical theory and performed numerical experiments to investigate the potential of working with a subset of the scenarios during robust optimization of IMPT. Three main aspects are considered:

The first part is dedicated to investigating the existence of an optimal subset of scenarios in the practical sense. Although the theory implies that there exists some subset of scenarios $\mathcal{A}^\star \subset \mathcal{S}$, such that any scenario in the complement $\mathcal{S} \setminus \mathcal{A}^\star$ is redundant from the perspective of solving Problem \eqref{eq_problem}, it is not implied that common methods for optimization of IMPT can reliably identify that subset. Furthermore, we are not aware of any guarantees whether the optimization trajectory associated with $\mathcal{A}^\star$ contains only points $\{ x^k_{\mathcal{A}^\star} \}^\infty_{k = 0}$ with a low robustness gap $\Delta (x^k_{\mathcal{A}^\star}, \mathcal{A}^\star)$. However, our experimental results indicate that it is indeed feasible to estimate such a subset, although it requires higher accuracy than is normal in radiation therapy optimization. In addition, they suggest that if such a set is known, it does indeed provide a low robustness gap not only near the optimal solution, but also earlier during the numerical optimization (after 50 iterations).

The second part concerns the extent to which simple, a priori methods can be used to replicate the robustness properties of $\mathcal{A}^\star$, which itself is not practical to compute. From a scenario set such as in the case of 4DRO, it may be possible to manually select, by intuition, the most important scenarios based on the geometry of the case. However, our results indicate that in some cases (2/6), this results in far worse objective values than when using the full set or the subsets found with the investigated diversity-maximization methods. Furthermore, these methods, or variations thereof, extend beyond the case of 4DRO in the sense that they can also be applied to other sources of uncertainty, for which it may not be straightforward to make a manual selection. For example, interplay-robust optimization has been proposed to mitigate uncertainty in the patient's breathing motion. There, the employed motion pattern scenarios are not easily discretized, nor easily ranked by difficulty. The same reasoning applies to robust optimization considering various ROI contour scenarios. To a large extent, this study is inspired by the \textit{relevant scenario recognition problem} presented in Goergik et al. \cite{goerigk_data-driven_2025}. There, the authors design a machine learning model to identify a relevant subset of scenarios, and train it on other instances of the problem. Although this approach is not necessarily practical for radiation therapy applications, it may be interesting to explore their method or other data-driven methods that intend to identify the relevant scenarios a priori. Our diversity-maximization approach is a first attempt at identifying features that determine a scenario subset with a low robustness gap.

Finally, the third part investigates the potential computational benefit of working with a reduced set of scenarios. In our experiments, considerable time was saved for all patients, with mean (across the investigated adversarial methods) speed-ups of 6.7--17.0x for five out of the six patients, and a mean speed-up of 2.8x for P111. For the five patients with greater gains from using the adversarial methods, later iterations of the benchmark method needed more time to solve the subproblems and were thus considerably slower than earlier iterations. This slowdown of the benchmark method considerably increased the gains of the adversarial methods, which did not suffer from the same drawback. In our experience, the time per iteration of SQP methods used to solve radiation therapy optimization problems formulated using sums of squared voxel-wise penalties tends to be dominated by the time taken to perform matrix-vector multiplications to compute the dose distribution(s) and the optimization function gradient(s). Without considering the nominal dose and objective, reducing the number of scenarios used proportionally reduces the number of times these operations are performed. In contrast, the time used to solve the QP subproblems is typically comparatively small. Therefore, we also considered the computation time savings earlier during the optimization (after 20 iterations), before the occurrence of the described slowdown of the benchmark. Then, the time savings were more modest but still considerable at 2.1-2.2x for P101 and P103 and 2.7-3.3 for P104, P110, P111, and P114. When solving for a greater number of iterations, we think that using a numerical method to solve subproblems that does not suffer from the encountered slowdown is likely to make the computational gains of the adversarial methods more similar to those recorded after 20 iterations.

This study does not consider anything beyond more efficiently solving Problem \eqref{eq_problem}. For example, we have not investigated the effects on commonly used evaluation metrics for radiation therapy plans, such as \textit{dose-volume histograms} (DVHs). Designing objectives that correlate well with clinically relevant metrics is a well-known challenge in radiation therapy treatment planning \cite{zhang_direct_2020}. To derive practical use out of the methods considered, a thorough investigation into their effects on DVHs and other clinically relevant metrics may be needed. Furthermore, we have not addressed the fact that for the investigated methods, the scenario-specific dose-influence matrices need to be stored in memory just as when solving Problem \eqref{eq_problem} directly. This is another challenge whose difficulty increases with the number of considered scenarios. Recently, \textit{beamlet-free} optimization and \textit{scenario-free} robust optimization have been proposed as alternatives that have the potential to address the increasing demands on both memory and computation \cite{pross_technical_2024, cristoforetti_scenario-free_2025}.

In conclusion, we have investigated aspects of working with a reduced scenario set during the robust optimization of IMPT plans. The results indicate that for the commonly used minimax formulation, the worst scenarios at the optimal solution are worst further from the optimum too, if used as a reduced scenario set in the robust optimization. In addition, efforts made to estimate this subset, either by diversity-maximization, adversarial methods, or a combination of both, reduced the computation time required to solve the minimax formulation by at least 50\%.

\printbibliography

\end{document}